\documentclass[sigconf]{acmart}

\usepackage{booktabs} 
\usepackage{enumitem}
\usepackage{subcaption}

\begin{document}
\title{Assessment of Prediction Techniques:\\ The Impact of Human Uncertainty}

\author{Kevin Jasberg}
\affiliation{%
  \institution{Web Science Group\\Faculty of Arts and Humanities}
  \city{Duesseldorf} 
  \state{Germany} 
  \postcode{40225}
}
\email{kevin.jasberg@uni-duesseldorf.de}

\author{Sergej Sizov}
\affiliation{%
  \institution{Web Science Group\\Faculty of Arts and Humanities}
  \city{Duesseldorf} 
  \state{Germany} 
  \postcode{40225}
}
\email{sizov@hhu.de}

\renewcommand{\shortauthors}{}

\pagenumbering{arabic}
\pagestyle{plain}
\thispagestyle{plain}
\begin{abstract}
Many data mining approaches aim at modelling and predicting human behaviour. An important quantity of interest is the quality of model-based predictions, e.g. for finding a competition winner with best prediction performance.

In real life, human beings meet their decisions with considerable uncertainty. 
Its assessment and resulting implications for statistically evident evaluation of predictive models are in the main focus of this contribution. We identify relevant sources of uncertainty as well as the limited ability of its accurate measurement, propose an uncertainty-aware methodology for more evident evaluations of data mining approaches, and discuss its implications for existing quality assessment strategies. Specifically, our approach switches from common point-paradigm to more appropriate distribution-paradigm.

This is exemplified in the context of recommender systems and their established metrics of prediction quality. The discussion is substantiated by comprehensive experiments with real users, large-scale simulations, and discussion of prior evaluation campaigns (i.a. Netflix Prize) in the light of human uncertainty aspects.
\vfill
\end{abstract}


\maketitle

\newpage

\section{Introduction}
A broad range of algorithms and approaches in data mining aims at modelling and predicting aspects of human behaviour. 
These efforts are motivated by many practically relevant applications, including various recommender systems, content personalisation, targeted advertising, along with many others. The comparative assessment of methods usually involves implicit or explicit knowledge about user behaviour, either by observing user interactions, or by asking users explicitly.

In many situations, particular individuals may meet their decisions with considerable uncertainty. In other words, they would not exactly reproduce their decisions when asked twice or multiple times. Consequently, observed decisions must be seen as single draws from individual ``feeling''-distributions, resulting from complex cognition processes, and influenced by multiple factors (e.g. mood, media literacy, etc.). Moreover, and even more important, our knowledge about such distributions may be very limited due to natural restrictions of human behaviour, i.e. it is practically not possible to require the necessary amount of repeated trials for precise location of the underlying distribution parameters.

The presence of human uncertainty and our incomplete knowledge about its properties naturally raise the question of assessment validity and reliability. If some approach $R1$ shows better results than approach $R2$ in the sense of a certain quality metric (prediction accuracy, user satisfaction, etc.) given reference data, can we consider this as a statistically evident proof that approach $R1$ \textit{is indeed} better? In the common sense of statistical hypothesis testing, the confident conclusion can be made if the opposite case has a very low probability (type I error) to happen. Under appropriate accounting for human uncertainty, such certainty is often hard to reach.

\paragraph{Motivating example}
As a motivating example, we consider the task of rating prediction 
(common to recommender systems research), along with the Root Mean Square Error (RMSE) \cite{Herlocker} as a widely used metric for prediction quality. 
In a systematic experiment with real users (described in more detail in the forthcoming sections), individuals rated certain media items (movie trailers) multiple times. 
Only 27\% of users have shown constant rating behaviour; 73\% of them have given at least two different ratings to the same item; 49\% of users have given three or more different responses. Based on the observations made so far, we constructed individual uncertainty models for every user and thus, the considered quality metric (in our case, RMSE) becomes a random variable which is distributed with respect to a certain probability density function.

\begin{figure}[t]
\centering
\includegraphics[width=.49\textwidth]{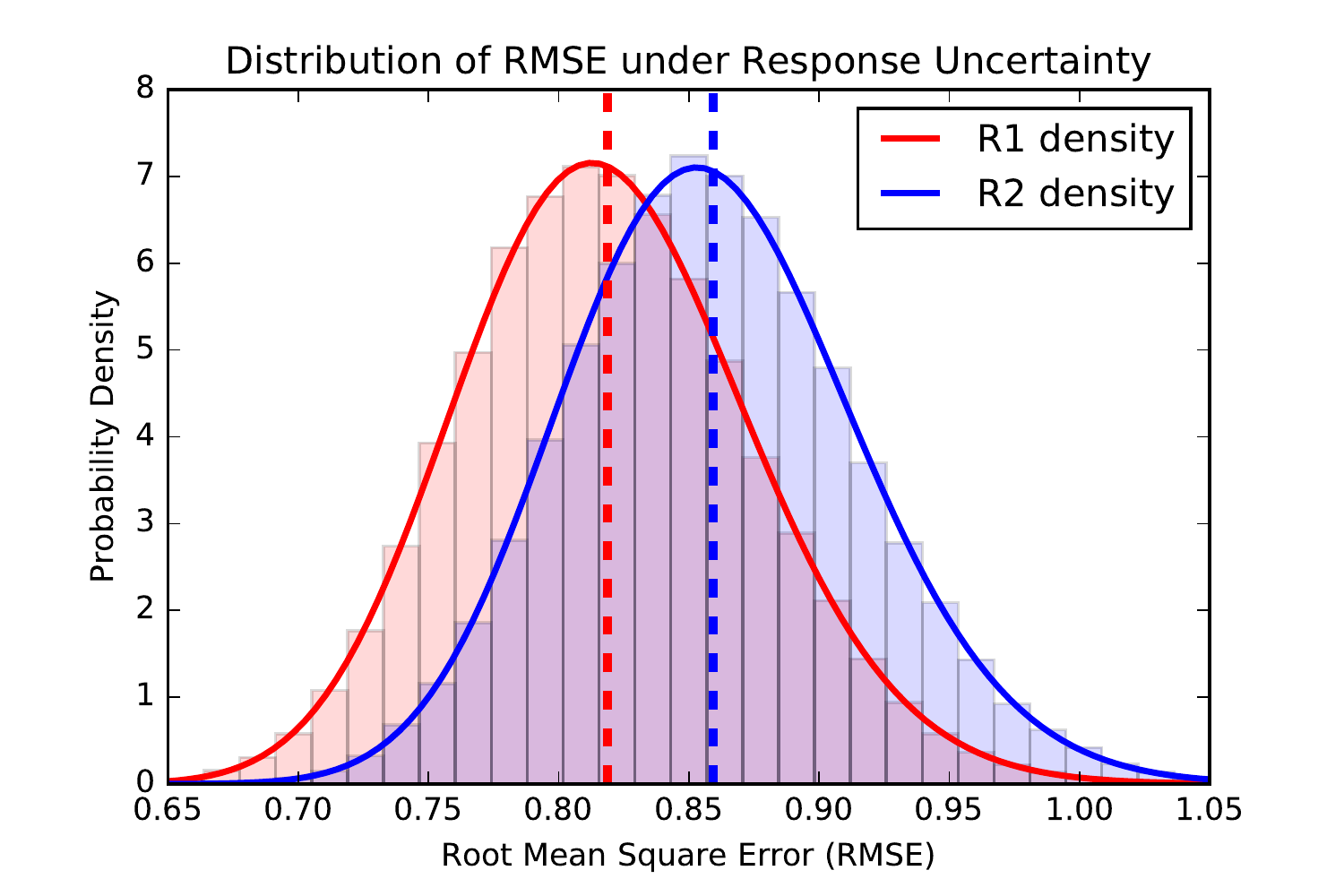}
\caption{Comparison of two recommender systems performing on a case study data set using the RMSE visualised in perspective of the point- and distribution-paradigm}
\label{fig:motivatingexample}
\end{figure}

Figure \ref{fig:motivatingexample} shows corresponding results for two sample recommenders that make a best possible prediction (the mean of observed user responses) (red chart) vs. random predictions around the mean (blue chart); R1 is supposed to be the better system by design. As can be seen, there is a large overlap between both PDFs inducing a probability of $P(\text{``}R2\text{ better than } R1\text{''})\approx 0.33$ that the worse recommender R2 can even outperform the superior R1. In other words, we would opt for the wrong recommender in $1$ of $3$ repetitions of the evaluation process when using single RMSE-scores rather than the entire distribution. Insofar, the simple comparison of point-wise calculated quality metrics is not necessarily evident for a statistically sound proof of method advantages. Without any loss of generality, the observations made so far can be considered as an indicative motivation for a more careful analysis of the following research questions:
\begin{description}
\item[Q1:] How well is human uncertainty measurable and what are the implications of its incomplete assessment onto possible model comparisons?
\item[Q2:] How well can distinguishability be reached under the human uncertainty assumption, specifically:
\begin{enumerate}[label=(\alph*)]
\item What is a natural metric for the distinguishability between two ``substantially'' different models? 
\item What kind of statistical evidence can indicate that a model can still be improved?
\item What makes a difference between two models statistically significant?
\end{enumerate}
\end{description}

\section{Related Work}

In the context of this paper, we exemplify our approach by scenarios from the field of recommender systems as summarised in \cite{handbook} and focus specifically on comparative evaluation metrics. Recommender systems were initially based on demographic, content-based and collaborative filtering. An overview of these techniques are given in \cite{Bobadilla}. As collaborative filtering recently turned out to be one of the most successful techniques, they rapidly got into the centre of further research. A roadmap to collaborative filtering as well as a profound discussion on its predictive performance is provided by \cite{Su}.

Due to the importance of evaluating those recommender systems in terms of their model-based prediction quality, different metrics have been introduced, such as the root mean squared error (RMSE), mean absolute error (MAE), mean average precision (MAP) and normalized discounted cumulative gain (NDCG) (see \cite{workshop12}). Further possible quality-related dimensions of interest in recommender assessment (user satisfaction, precision/recall, etc.) are summarised in \cite{Herlocker}. 

All mentioned quantities have in common the need for human input, either by asking the users explicitly, or by observing their interactions. In both the cases, human responses may show a considerable degree of uncertainty, resulting from complex cognition processes and multiple influential factors. Consequently, the main results shown in this contribution can be easily adopted for general cases without substantial loss of validity.

The idea of uncertainty is not only related to predictive data mining but also to measuring sciences such as physics or biology. In this area, a science called metrology has been developed, which is about accurate and precise measurement. Recently, a paradigm shift was initiated on the basis of a so far incomplete theory of error (see \cite{Grabe}), so that variables are currently modelled by probability density functions and quantities calculated therefrom are now assigned a distribution by means of a convolution of these densities. This model is described in \cite{GUM}. A feasible framework for computing these convolutions via Monte-Carlo-simulation is given by \cite{GUMsupp1}. We take this model as a basis for our modelling of uncertainty for addressing similar issues in the field of computer science.

The complexity of human perception and cognition can be addressed by means of latent distributions (see \cite{CogPsych}), resulting in varying observations. This idea is widely used in cognitive science and in statistical models for ordinal data. For example, so-called CUB models for ordinal data \cite{cub} assume the Gaussian as a latent response model underlying the observations. We adopt the idea of modelling user uncertainty by means of individual Gaussians following the argumentation in \cite{cub} for constructing our own response models. 

The human impact on the prediction quality was noticed in 2009 when \cite{LikeLikeNot} stated, that users are inconsistent in giving feedback and therefore establish an unknown amount of noise that challenges the validity of collaborative filtering. In consequence, further improvements of prediction accuracy, that don't particularly consider user noise, have been proven obsolete by \cite{MagicBarrier}, i.e. the human impact leads to a natural non-vanishing offset for any metric that we cannot overcome.

In order to collect information about human uncertainty, we follow \cite{RateAgain} by using repeated rating scenarios for same users and items within conducted experiments designed in accordance with experimental psychology \cite{psycho1, psycho2}. 
On the basis of the informations gathered by using this approach, the authors of \cite{RateAgain} were able to develop a pre-processing in order to de-noise the underlying data set of ratings and therefore yield better prediction accuracy. In contrast we distinguish between non-significant deviations (natural human noise) and significant ones (model induced noise). In this paper, we use the same measuring instrument to collect uncertainty information as in \cite{RateAgain} but in this contribution, we also focus on the influence of this uncertainty on the accuracy of recommender systems under the view of metrology. We also take the idea of a pre-processing to reduce the impact of human uncertainty on RMSE under this different perspective.

\section{Modelling Human Uncertainty}
For evaluating the quality of model-based predictions exemplified by recommender system accuracy, we compare internally computed predictors against real user ratings. Let $\mathcal{I} = \{ 1,\ldots, I \}$ be the index set of $I$ items and $\mathcal{U}= \{1,\ldots, U \}$ the index set of $U$ users. When several users have rated several items, we obtain $n\leq U\cdot I$ pairs $(\pi_\nu  , r_\nu)$ of predictors $\pi_\nu $ and ratings $r_\nu $ that can be matched against each other where $\nu \in\mathcal{U}\times\mathcal{I}$ is a multi-index. These quantities allow to compute single scores of accuracy metrics (e.g. RMSE) which corresponds to the commonly used point-paradigm. By using the metrologic distribution-paradigm instead, we explicitly account for human uncertainty and its resulting rating uncertainty.

We consider all the given ratings to be a family of random variables $R_\nu  \sim\mathcal{N}(\mu_\nu,\sigma_\nu)$ which are assumed to be normally distributed as also done in \cite{cub}. From this point of view, a given rating $r_\nu$ can be seen as the output of a random experiment that is somehow related to human cognition. Hereunder, human uncertainty is strongly related to statistical randomness and the standard deviation $\sigma_\nu$ becomes a natural measure of human uncertainty.
In this case, the RMSE becomes a random variable itself since it is a composition of continuous maps of random variables. The distribution emerges as a convolution of $n$ density functions under the given mathematical model
\begin{equation} \label{eq:RMSE}
\text{RMSE} = \sqrt{\sum_{\nu\,\in\,\mathcal{U}\times\mathcal{I}} \frac{(\pi_\nu  - R_\nu )^2}{n}}.
\end{equation}

As an example, we consider all $n$ rating distributions to be i.i.d. with $R_\nu\sim\mathcal{N}(\pi_\nu,1)$ that is, the predictors of our recommender systems perfectly match with the mean of our rating distributions. With these distributions we want to derive the RMSE's density gradually by specifying the densities for every step of computation that has to be done for calculating the entire RMSE. First we consider the initial step 
$S^1_\nu := \pi_\nu - R_\nu$ which is a random variable distributed by $\mathcal{N}(0,1)$.
Then as sum of $n$ standard normal distributed random variables, the second step $S^2_\nu := \sum_\nu (S^1_\nu)^2$ yields a $\chi^2(n)$-distribution with $n$ degrees of freedom. Hence, a scaling by $1/n$ will lead to a gamma distribution
$S^3_\nu :=  \tfrac{1}{n} \cdot S^2_\nu \sim \Gamma(\tfrac{n}{2},\tfrac{2}{n})$ and finally for the last step, 
$S^4_\nu := \sqrt{Z^2_\nu} \sim \text{Nakagami}(\tfrac{n}{2},1)$ yields the Nakagami-distribution since it is the square root of a gamma-distributed random variable. Under all these conditions, we yield the RMSE not to be a single point but rather to be a $\text{Nakagami}$-distributed random variable with density function
\begin{equation} \label{eq:Nakagami}
f(x) = \frac{2m^m}{\Gamma(m)}x^{2m-1}\exp\left(-mx^2\right) \quad\text{where}\quad m=n/2.
\end{equation}
whose expectation
\begin{equation} \label{eq:RMSEexpectation}
\mathbb{E}(\text{RMSE})=  \frac{\Gamma(\tfrac{n+1}{2})}{\Gamma(\tfrac{n}{2})} \sqrt{\frac{2}{n}}
\end{equation}
is the average RMSE score according to the point paradigm when repeating the rating scenario infinitely. The advantage of this approach is, that it additionally provides a non-vanishing variance 
\begin{equation} \label{eq:RMSEvariance}
\mathbb{V}(\text{RMSE}) = 1-\frac{2}{n}  \cdot \left(   \frac{\Gamma(\tfrac{n+1}{2})}{\Gamma(\tfrac{n}{2})}\right)^2
\end{equation}
as a measure for the uncertainty that is related the RMSE. The fact that a different RMSE score is achieved each time the rating scenario is repeated, corresponds to drawing a random number from a given RMSE distribution within the distribution-paradigm.
Considering a data set of uncertain ratings, two different recommender systems would gain different RMSEs on this dataset,  denoted $X_1$ and $X_2$ .
Let $f_{X_1}(x)$ and $f_{X_2}(x)$ the probability density functions of $X_1$ and $X_2$. If those densities overlap, i.e. the quantity 
\begin{equation}
A:=\int_{-\infty}^\infty \vert f_{X_2}(x) - f_{X_1}(x) \vert \,\mathrm{d}x
\end{equation}
does not vanish, then there is also a non-vanishing possibility of error when building a ranking order by evaluating single scores only (point-paradigm). Let $x_1$ and $x_2$ denote two realisations of the RMSEs $X_1$ and $X_2$ and let $x_1 < x_2$ be the ranking order by using the point-paradigm, then the probability $P_\varepsilon $ of error for this decision is given by $P_\varepsilon := P(X_1>X_2)$ with
\begin{equation}\label{eq:ErrorProb}
P(X_1>X_2) := \int_{-\infty}^\infty f_{X_2}(x) \big( 1-F_{X_1}(x) \big) \,\mathrm{d}x \leq 0.5
\end{equation}
where $F_{X_1}(x):=\int_{-\infty}^x f_{X_1}(t)\,\mathrm{d}t$ denotes the cumulative distribution function of $f_{X_1}$.
Later, it will be shown that a ranking built by using the point paradigm is associated with considerable errors caused by human uncertainty. However, this can virtually be subtracted out by a pre-processing step. 

From the view of the distribution-paradigm, each time a given rating is compared with a model-based prediction,
we must examine whether the observed deviations are significant or just in nature of contingency, i.e. the influence of human uncertainty. In other words, we must divide the set of all deviations into two subsets. One subset contains all the deviations around the predictor $\pi_\nu$ that can be considered as human uncertainty and the other subset contains all deviations whose extent cannot be explained by this uncertainty and thus seems to be induced by the predictor model. In this case it seems viable to calculate the RMSE by taking into account only those deviations  that are related to the algorithm rather than to human uncertainty. Similarly to the classic RMSE we refer to this more natural metric as the significant RMSE (\textbf{sRMSE}). Following this approach we have to use statistical hypothesis testing to decide whether a realisation $ r_\nu$ of the rating distribution $R_\nu$ is equal to a model-based prediction $\pi_\nu$ or not. In mathematical notation, we have to test
\begin{equation}
H_0 \colon  r_\nu = \pi_\nu \quad\textbf{vs.}\quad H_1 \colon  r_\nu \neq \pi_\nu
\end{equation}
for every multi-index $\nu$ at a given significance level $\alpha$.
For known density functions $f_{R_\nu}$ of the rating distributions $R_\nu$ the critical region can be constructed as the complement of $I_{\alpha} = [ \pi_\nu - a  \,;\, \pi_\nu + a]$ where $a$ is chosen such that
\begin{equation}\label{eq:confinterval}
\int_{\pi_\nu - a}^{\pi_\nu + a} f_{R_\nu}(x) \,\mathrm{d}x = 1-\alpha.
\end{equation}
We now yield the probability density function of the sRMSE by a convolution of the pseudo-restrictions
\begin{equation}
f_{R_\nu}|_{I_{95}^\complement}(x) := \mathbb{I}_{I_{95}^\complement}(x) \cdot f_{R_\nu}(x)
\end{equation}
where $\mathbb{I}$ is the indicator function.
Due to this definition, the sRMSE grants assessment of different recommender systems with much lower probabilities of error. This can be explained by not taking into account the stabilising centre of all the rating-distributions and as the RMSE amplifies the remaining extreme values by its quadratic term (see Equation \ref{eq:RMSE}), the distributions rapidly differ under increasement of false predictions. 
Having in mind this mathematical model of human uncertainty in terms of the novel metrologic distribution-paradigm, we elaborate on our research questions by examination of real life scenarios.

\section{User Study and Simulations}
In practice, the application of the previously described model is technically challenging.
Let the rating distributions $R_\nu\sim\mathcal{N}(\mu_\nu,\sigma_\nu)$ be not necessarily equal for every $\nu$. 
As it has been shown in \cite{Chan}, the sum of squared deviations receives the density of a non-central 
$\chi^2$-distribution. At this point it is quite hard to find a closed form for the RMSE density. It turns out that efficient dealing with the RMSE's distribution can only be maintained by using statistical simulations when general cases are taken into account.
In this paper we use \textbf{Monte-Carlo-Simulations} (MC) as described in \cite{GUMsupp1}: 
For every input variable $R_\nu\sim\mathcal{N}(\mu_\nu,\sigma_\nu)$ we take a sample 
$\mathcal{S}(R_\nu):= \{ r^1_\nu,\ldots, r^\tau_\nu\}$ of $\tau$ pseudo-random numbers (trials) that are drawn from this specific distribution. 
Due to the randomness further computations may fluctuate slightly, but his effect diminishes for a high number of trials. In our analyses we reached stable results by setting $\tau=10^6$. 
With these samples we compute $\mathcal{S}(\text{RMSE})$ by
\begin{equation}\label{eq:RMSE_MCM}
\mathcal{S}(\text{RMSE}) =
\left\lbrace 
y_j = \sqrt{\sum_{\nu} \tfrac{(\pi_\nu  - r^j_\nu )^2}{n}} \,\colon j=1,\ldots,\tau
\right\rbrace.
\end{equation} 
Post hoc illustration of this sample by a normalised relative histogram with $b$ bins lead to an approximation of the RMSE's density. To be more precise: The envelope $E(x)$ of this histogram will converge to the true probability density function of the RMSE via
\begin{equation} \label{eq:DensityConvergence}
f_{RMSE}(x) = \lim_{\substack{b\to\infty \\ \tau\to\infty}} E(x).
\end{equation}

Our analyses often focus on the error probability $P_\varepsilon$ as described in Equation \ref{eq:ErrorProb}.
In the following numerical simulations this probability is efficiently computed by
\begin{equation} \label{eq:NumError}
P_\varepsilon= P(\text{RMSE1}>\text{RMSE2}) = \vert A\vert / \tau
\end{equation}
where $A$ is the set of all $(r_i,s_j)\in\mathcal{S}(\text{RMSE1})\times \mathcal{S}(\text{RMSE2})$
holding the condition $r_i >s_i$ for $i=1,\ldots,\tau$.

For modelling human uncertainty we assume a set of known rating distributions.
For the upcoming simulations, we estimated rating distributions based on perceptions about real user behaviour from comprehensive user experiments.

\subsection*{User Experiments}
Our experiment is set up with Unipark's\footnote{http://www.unipark.com/de/} survey engine whilst our
participants were committed by the crowdsourcing platform Clickworker\footnote{https://www.clickworker.de/}.
During the experiment, participants watched theatrical trailers of popular movies and television shows and provided ratings on a 5-Star-Likert-Scale multiple times in random order. The submitted ratings have been recorded for five out of ten fixed trailers so that the remaining trailers act as distractors triggering the misinformation effect, i.e. memory is becoming less accurate because of interference from post-event information.

The experiment starts with an introductory phase in which four very short trailers are shown and
rated. One of these introductory trailers is shown twice to prepare the participants for an upcoming redundancy so that no biasing confusion arises in the further progress. 

During the initiation of the main phase in which we start recording the user ratings for five predetermined
trailers, every trailer is shown once and has to be seen completely before giving a rating. 
Afterwards, ratings can be submitted after 20 seconds to ensure a shortening of imposed runtime and prevention of rapid loss of interest when watching the same trailer multiple times. We also supported this intention by adding five additional trailers randomly which are not to be repeated, in order to maintain a user's interest. However, this provides a positive side effect: It also prevents the users to start rating repeated trailers in relation to each other which is likely to occur when displaying the same items in different orders too often.

Altogether, we received a Rating-Tensor $R_{u,i,t}$ with $\dim(R)=(67,5,5)$, having $N=1675$ data points in total, where the coordinates $(u,i,t)$ encode the rating that has been given to item $i$ by user $u$ in the $t$-th trial. From this dataset we derive a unique rating distribution for every user-item-pair by considering tensor-slices in time-dimension 
$R_{u,i}:= R_{u,i,\bullet}= \{ R_{u,i,t} \vert t=1,\ldots ,5 \}$ which can be easily depicted in a relative histogram and modelled by a certain rating distribution. 

From our experiment, only a few tensor slices contain constant ratings and hence lead to a vanishing variance.
\begin{table}[t]
\centering
\renewcommand{\arraystretch}{1.2}
\begin{small}
\begin{tabular}{|c|c|c|}         
\cline{2-3}
\multicolumn{1}{c|}{}   &       $\operatorname{var}\neq 0$      &       normality not rejected\\ \cline{1-3}
item 1                                  &       0.90    &       1.00                                    \\
item 2                                  &       0.60    &       1.00                                    \\
item 3                                  &       0.50    &       1.00                            \\
item 4                                  &       0.69    &       1.00                                    \\
item 5                                  &       0.51    &       1.00                                    \\ \cline{1-3}
\end{tabular}
\end{small}
\caption{Fraction of noisy data and not rejected normality for the re-rating-proceeding (relative frequencies)}
\label{tab:ExpRes}
\end{table}
As we can see in Table \ref{tab:ExpRes}, the fraction of tensor slices with non-zero variance is ranged from 50 to 90\% that is, only every second participant is able to reproduce its own decisions for the best case. In the worst case only one out of ten participants is able to precisely reproduce a rating. All tensor slices containing a non-vanishing variance are checked for normality by a KS-test at $\alpha=0.05$. As a result, the null hypothesis is never rejected, retaining the normal distribution to be a possible model since none of these samples actually differs from it significantly.

\subsection*{Research Question Q1: Measurability of Human Uncertainty and Implications}
\textbf{Description:}
Based on our user study, we assume $R_\nu\sim\mathcal{N}(\mu_\nu,\sigma_\nu)$.
Since this study only surveyed a sample rather than an entire population, point estimates for the distribution parameters would be inappropriate. Instead, confidence intervals have to be specified. Following \cite{MuConfInt}, the confidence interval for the parameter $\mu_\nu$ can be received by
\begin{equation}\label{eq:CIm}
\mu_\nu\in 
\left[
\bar{x}_\nu - t_{(1-\tfrac{\alpha }{2};n-1)}  \frac {s_\nu}{\sqrt {n}}
\ ;\ 
\bar{x}_\nu + t_{(1-\tfrac{\alpha }{2};n-1)}  \frac {s_\nu}{\sqrt {n}}
\right]
\end{equation}
where $\bar{x}$ and $s$ are the point estimates for the mean and bessel-corrected standard deviation and
$t_{(p;k)}$ represents the $p$-quantile of the t-distribution with $k$ degrees of freedom. Following \cite{VarConfInt}, the confidence interval of $\sigma_\nu$ is given by
\begin{equation} \label{eq:CIs}
\sigma \in \left[
s{\sqrt {(n-1)/\chi _{(1-{\tfrac {\alpha }{2}};n-1)}^{2}}}
\ ; \ 
s{\sqrt {(n-1)/\chi _{({\tfrac {\alpha }{2}};n-1)}^{2}}}
\right]
\end{equation}
where $\chi _{(p;k)}^{2}$ is the $p$-quantile of the $\chi^2$-distribution with $k$ degrees of freedom.
This means that we can not simply determine a single rating distribution for each data set.
Instead, a variety of rating distributions need to be computed for each user-item-pair where the 
associated parameters are drawn from the corresponding confidence interval. Even for large-scale computations the resulting RMSE does not possess a stable density function. However, we can consider borderline cases which reveal the maximum span in which we can expect results for the density function of the RMSE.

\noindent
On this basis we run three simulations:
\begin{description}
\item[Simulation 1:] In Simulation 1 we compute these borderline cases by assigning the parameters $\mu_\nu$ and $\sigma_\nu$ as the lower limits of the corresponding confidence interval and the upper limits respectively. In doing so, we first build six recommender systems by defining their predictors via 
\begin{equation}
\pi^k_{(u,i)}:=
\begin{cases}
1/n\cdot \sum_t R_{u,i,t} & k=1 \\
 R_{u,i,k} & k=2,\ldots,6
\end{cases}
\end{equation}
where $k$ denotes the $k$-th recommender systems. Then, for every recommender systems we compute a sample $\mathcal{S}(\text{RMSE}(R\,k))$ for all borderline cases as described in Equation \ref{eq:RMSE_MCM} and generate the ML-density functions. In this simulation we use 
$\tau=10^6$ MC-trials for steadiness of samples' histograms as well as $b=55$ bins for accurate display of densities.

\begin{figure}[t]
    \centering
    \begin{subfigure}{0.49\textwidth}
        \includegraphics[width=\textwidth]{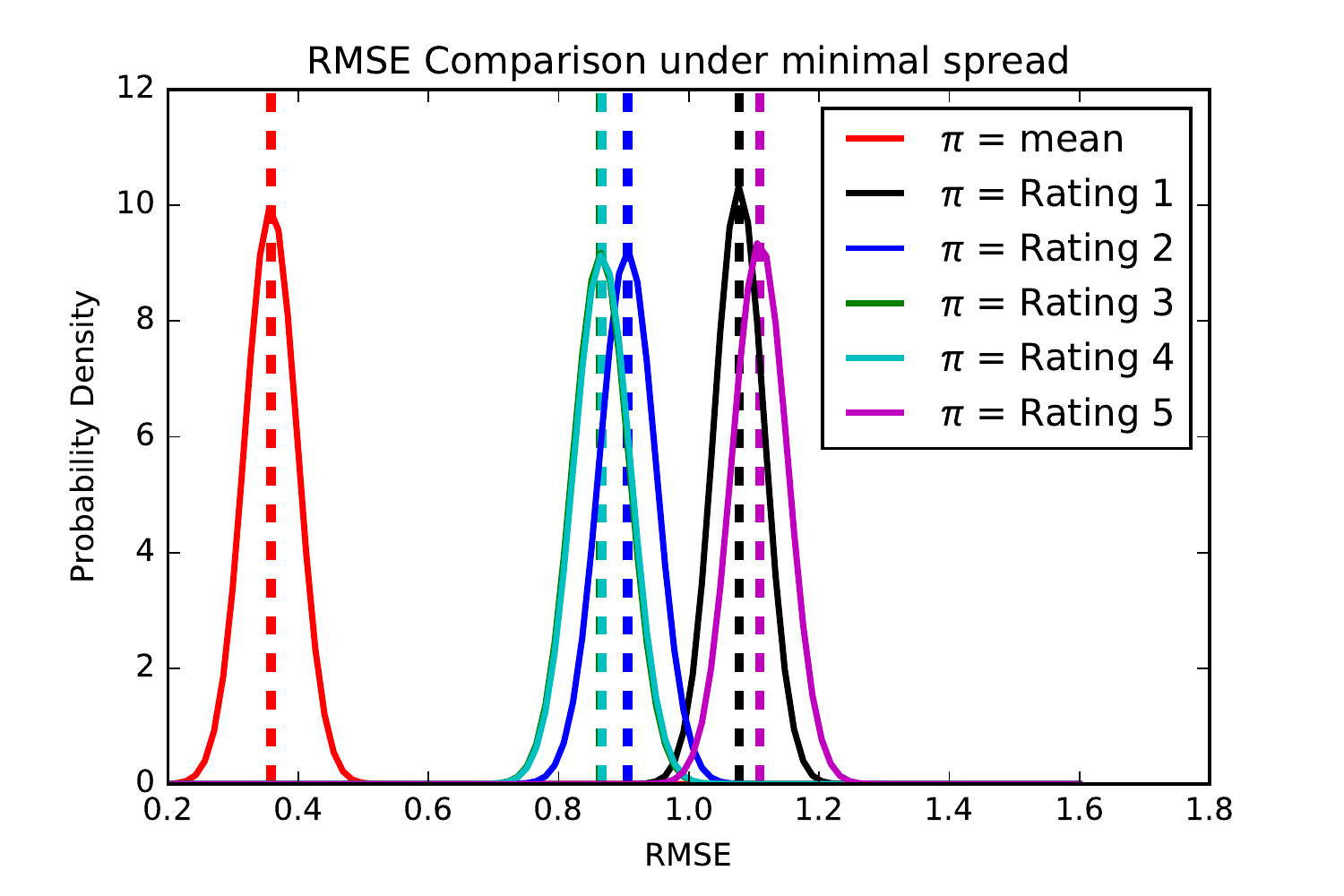}
        \caption{minimum case}
    \end{subfigure}
    \begin{subfigure}{0.49\textwidth}
        \includegraphics[width=\textwidth]{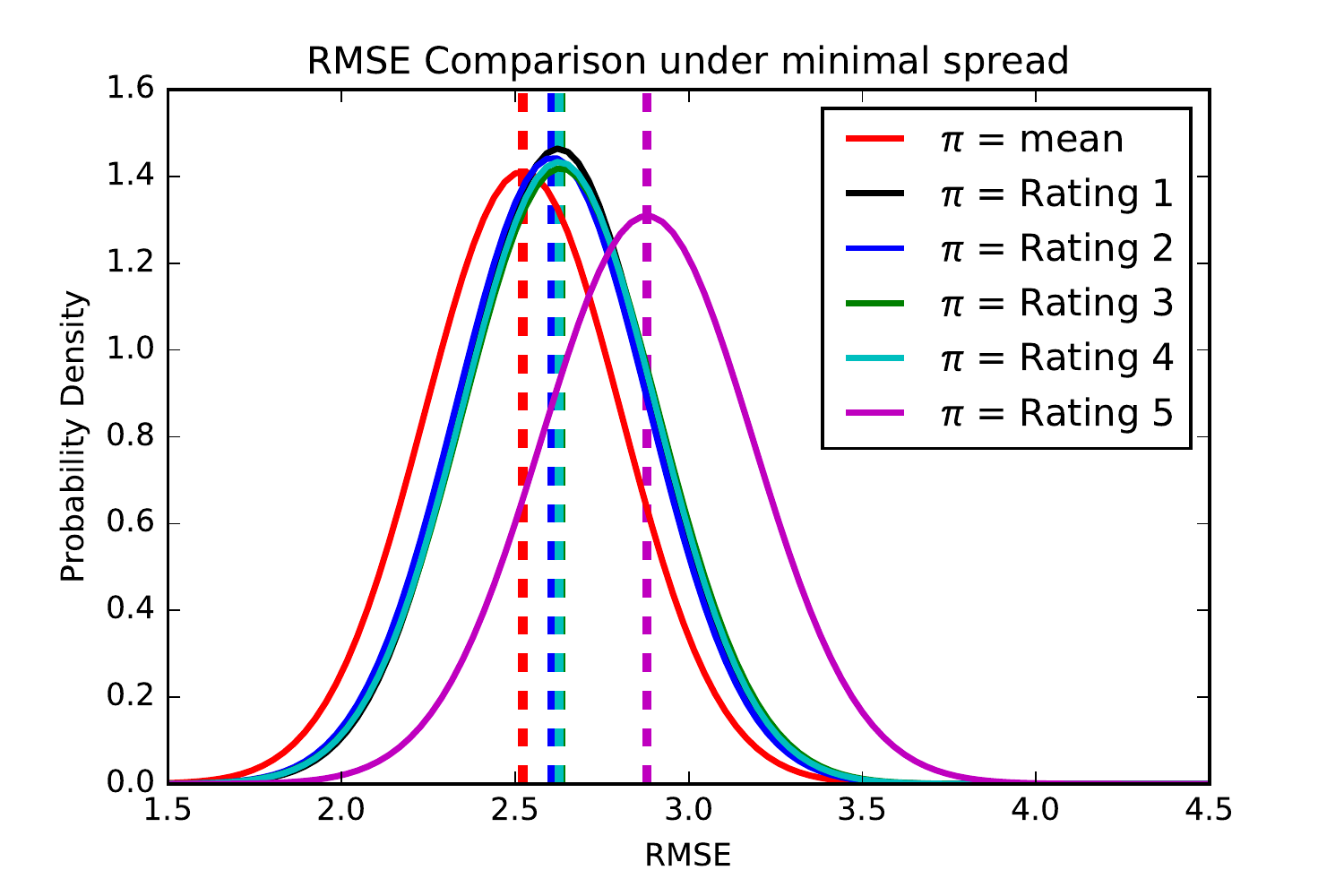}
        \caption{maximum case}
    \end{subfigure}
    \caption{Borderline cases of the RSMEs of different RS}
    \label{fig:MinMaxRMSE}
\end{figure}

Figure \ref{fig:MinMaxRMSE} shows the impact of the uncertainty of the re-rating-proceeding. Whilst we can recognise a good resolution for three groups of RMSEs in the minimum case, this is virtually no longer possible for the maximum case.
The true distributions of the individual RMSEs can vary between these two thresholds, but remain unknown to us on the basis of the information collected. In short, with only five re-ratings it is not possible to get high quality uncertainty information, but it must be said that this phenomenon is not grounded within the point-paradigm itself. In practice, we have to distinguish between two different types of uncertainty: On the one hand, there is the human uncertainty (leading from scores to distributions) which is in the main focus of this contribution. But on the other hand, there is also a kind of measurement error which we call the method uncertainty. The variability for the RMSE distributions in Figure \ref{fig:MinMaxRMSE} is completely explained by the impact of this method uncertainty.

\item[Simulation 2:] The method uncertainty can be reduced by increasing the number of re-ratings.
For this purpose, it is necessary to reduce the width of the confidence intervals that scale with $1/n^q$ for some 
$q\in\mathbb{R}$. Thus, the larger our sample of re-ratings, the smaller the intervals, i.e. the thresholds converge to the expected value for the respective parameter $\mu_\nu$ or $\sigma_\nu$. Accordingly, the borderline cases of the RMSE converge to a stationary state for large $n$. In this Simulation we estimate the amount of re-ratings to get stable results, so we can speak of \textit{true} RMSE of a recommender systems. As a measure of this convergence, we calculate the intersection area of the minimum and maximum RMSE for each recommender systems.

As can be seen from Figure \ref{fig:Intersection}, we need about 1000-2000 re-ratings, so that both distributions converge to a steady state by more than 90\%. This means that users in a real rating scenario would have to re-evaluate the same item at least 1000 times in order to locate the RMSE-distribution accurately.
\begin{figure}[b]
\centering
\includegraphics[width=.49\textwidth]{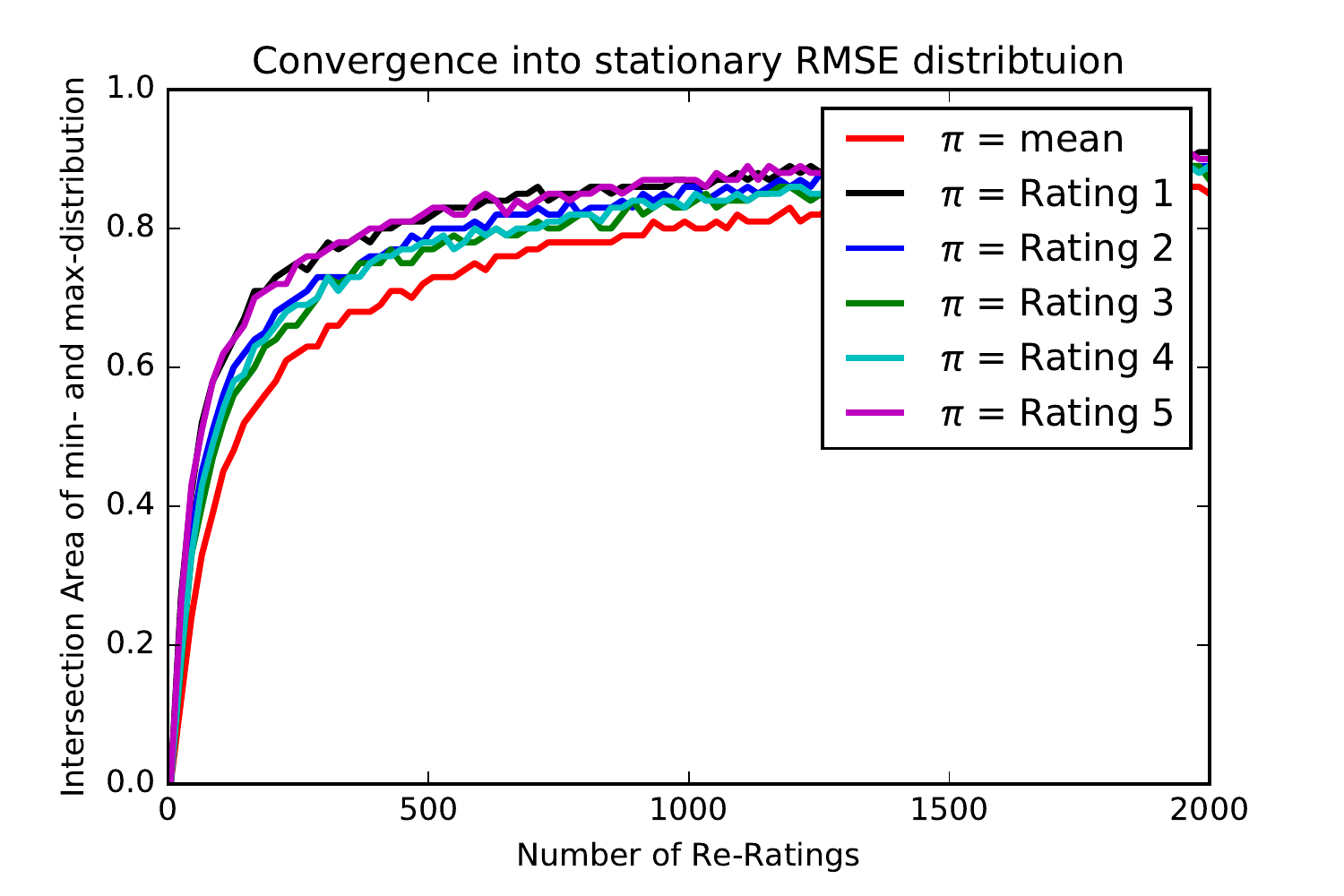}
\caption{Convergence of the minimum and maximum RMSE into a stationary state by means of their intersection}
\label{fig:Intersection}
\end{figure}

\item[Simulation 3:] If it is not feasible to calculate the stationary state with the re-rating-proceeding, then it might be sufficient to only gather samples as large as to exclude the high error probabilities of the maximum case. This is simulated by fixing the point estimates $\bar{x}$ and $s$ and artificially increasing the sample size $n$ to calculate the boundary points of our confidence intervals in Equation \ref{eq:CIm} and \ref{eq:CIs}. With those we determine the error probabilities for a point-paradigm ranking of recommender system 1 to all the other recommender systems for each of the borderline cases as described in \ref{eq:DensityConvergence}. 
Figure \ref{fig:ErrorConvergence} depicts the error probabilities $P_\varepsilon=P(\text{RMSE}(R1) > \text{RMSE}(R3))$ for the minimum and the maximum case. All the other cases of $P_\varepsilon=P(\text{RMSE}(R1) > \text{RMSE}(R\,k))$ lead to equivalent results for $k\neq 1$. As we can see, we would need about 500 re-ratings to regard the RMSE approximation to be satisfactory, if we accept a maximum of $P_\varepsilon\approx 0.10$. 
\begin{figure}[t]
\centering
\includegraphics[width=.9\linewidth]{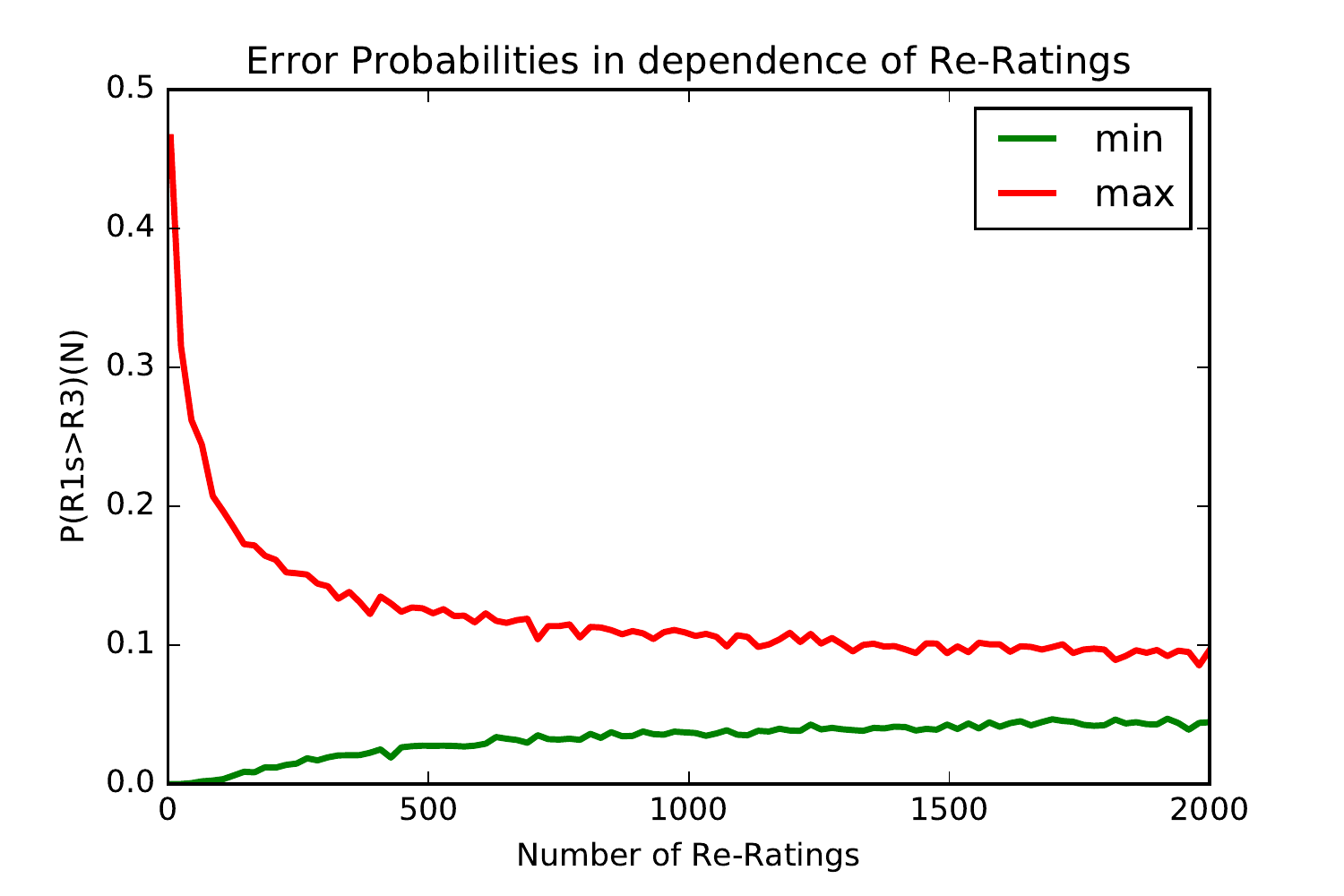}
\caption{Convergence of the error probabilities for a point-paradigm oriented ranking for the minimum and maximum RMSE distributions of RS 1 and 3}
\label{fig:ErrorConvergence}
\end{figure}
\end{description}

In some additional experiment we were able to show that the re-rating-proceeding loses validity for more than five trials.
For this purpose, we required 110 individuals to rate 220 stimuli chosen from photographs offered by Flickr\footnote{http://www.flickr.com} under Creative Commons licence. The photos show attractions of major European capitals (such as Eiffel Tower in Paris, Brandenburger Tor in Germany, etc.) from different perspectives, under different light conditions (day/night) and at different distances. Every attraction is present in the set a multiple times (4-6), in order to reduce the probability that respondents remember their opinion on a particular photo scene. Five of these stimulis were repeatedly rated whereas the other pictures served as distractors forcing the respondents to restart cognition for each picture rating. For any of the five photographs, we constructed individual rating-distributions under normality assumption and compared their mean and variance within every cohort (e.g. variances of all rating distributions for photograph 1 against the variances of all rating distributions for photograph 2, etc.). In doing so, we used Welch's t-test for comparing the mean as well as Levene's test for comparing the variances. Results prove that already after the first trial, a significant shift occurs in the expected values of the rating-distributions, whereas the variances remain constant. This phenomenom may be related to cognitions of learning and exploring the stimuli as well as the survey interface.
Trials two to four are stable therefrom, i.e. there is no significant change of the mean or variance respectively. These trials form the realm within the instrument of re-rating seems to measure with validity. In the fifth rating trial, the variances increase significantly, which can be explained by fatigue of the respondents, i.e. after rating the same stimuli four times, individuals will start rating randomly rather than deliberately. In brief, the re-rating is subject to strong natural limitations emerging from human behaviour. Together with the results from simulation 2 and 3, we can therefore state that the method of re-rating-proceeding described in \cite{RateAgain} is not able to measure the human uncertainty with sufficient accuracy and hence, precise statements about the true RMSE are not possible; a high overlap of two RMSE densities together with high error probabilities of a ranking according to the point-paradigm can never be excluded within prediction quality assessment.

\subsection*{Research Question Q2b: Statistical Evidence for Further Improvements}
Here, we examine the conditions under which a single recommender system can not be distinguished from a theoretically optimal recommender system by means of the RMSE. 
The idea of this investigation is to create a copy of a given recommender system and to distort this copy by artificial uniform-noise. 
This is done by resampling its predictors $\pi_1 \in [(1-p) \pi_0 \,;\, (1 + p) \pi_0] $ assuming a uniform distribution. 
In this case, a noise fraction of $p$ means that those new predictors deviate from the originals by $100p$\%. 
The RMSE thereby receives a shift on the x-axis so that it's possible to calculate a ranking including error probability. We can apply these as a function of the noise component. Noise is, in this context, a specific quantity for inducing differences in recommender system quality in a controlled manner. 

\begin{description}
\item[Simulation 4:] The expected value of a random variable is the value which is obtained on average in the case of an infinite repetition of the random experiment and thus has the smallest sum of squared deviations. Theoretically, this property makes the arithmetic mean $\bar{x}_{u,i}$ of the data series $R_{u,i}$ the optimal predictor. Hence, we define the optimal recommender system by setting 
$\pi_{u,i}:=\bar{x}_{u,i}$, so statements can be generated which are correct for very large investigations on the average. To this optimum we additionally create a copy which we distort by artificial uniform-noise as described and specify that two recommender systems can be distinguished significantly, if the error probability is less than 5\%. In this simulation we again use $\tau=10^6$ MC-trials for each of the $10^6$ data points $(p,P_\varepsilon)$, having $10^{12}$ trials in total.

\begin{figure}[b]
\centering
\includegraphics[width=0.49\textwidth]{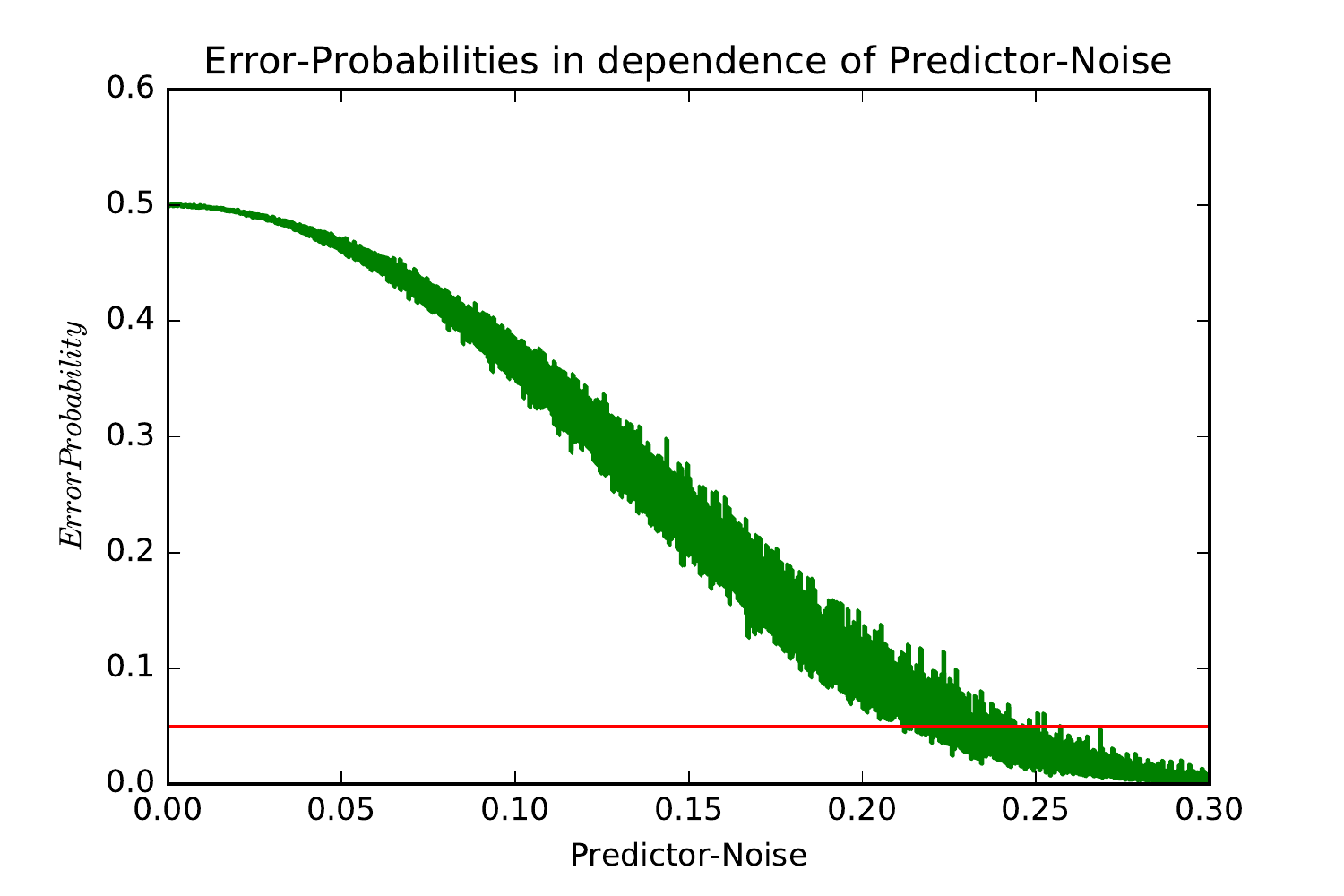}
\caption{Error probability for the point-paradigm ranking as a function of artificial predictor noise}
\label{fig:OptNoiseRMSE}
\end{figure}
Figure \ref{fig:OptNoiseRMSE} shows the curve of the error probability where the width of this graph is an artefact of the uniform-noise. We can see that the error probability drops below the 5\% mark in a range of 21\% to 24\%, i.e. only then distinctions to the optimum can be reliably detected. This proves the existence of a certain borderline of prediction quality so that any superior recommender system can not be differentiated from the best possible recommender system anymore.
\end{description}

\subsection*{Research Question Q2c: Significant Differences of two Models}
In real life, assessments compare several recommender systems among each other. This is taken into account in the following simulations. 
\begin{description}
\item[Simulation 5:] We generate two copies of an optimal recommender, 
with different proportions of added noise in such a way that the relative noise difference of both copies remain constant. Then, we compute the resulting RMSEs for both copies together with an error probability for the point-paradigm ranking. By increasing the noise for both copies whilst keeping their relative difference constant, we generate an offset (deviation from the optimum or prediction quality) and can thus apply the error probabilities against this offset for different noise ratios. This simulation was performed with $10^{12}$ data points.

\begin{figure}[t]
    \centering
    \begin{subfigure}{0.49\textwidth}
        \includegraphics[width=\textwidth]{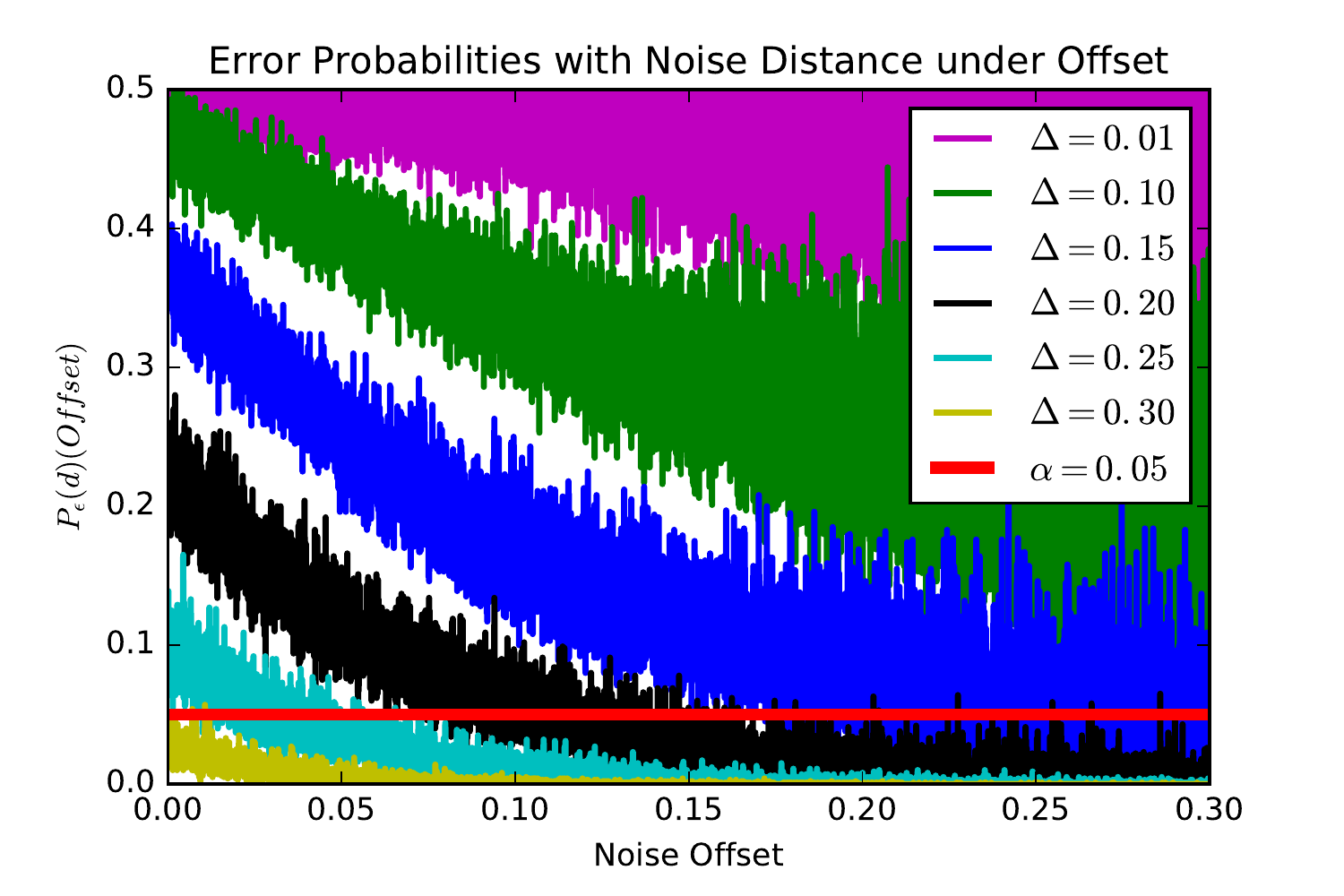}
        \caption{using noise as measure of distinction}
        \label{fig:NoiseDist}
    \end{subfigure}
    \begin{subfigure}{0.49\textwidth}
        \includegraphics[width=\textwidth]{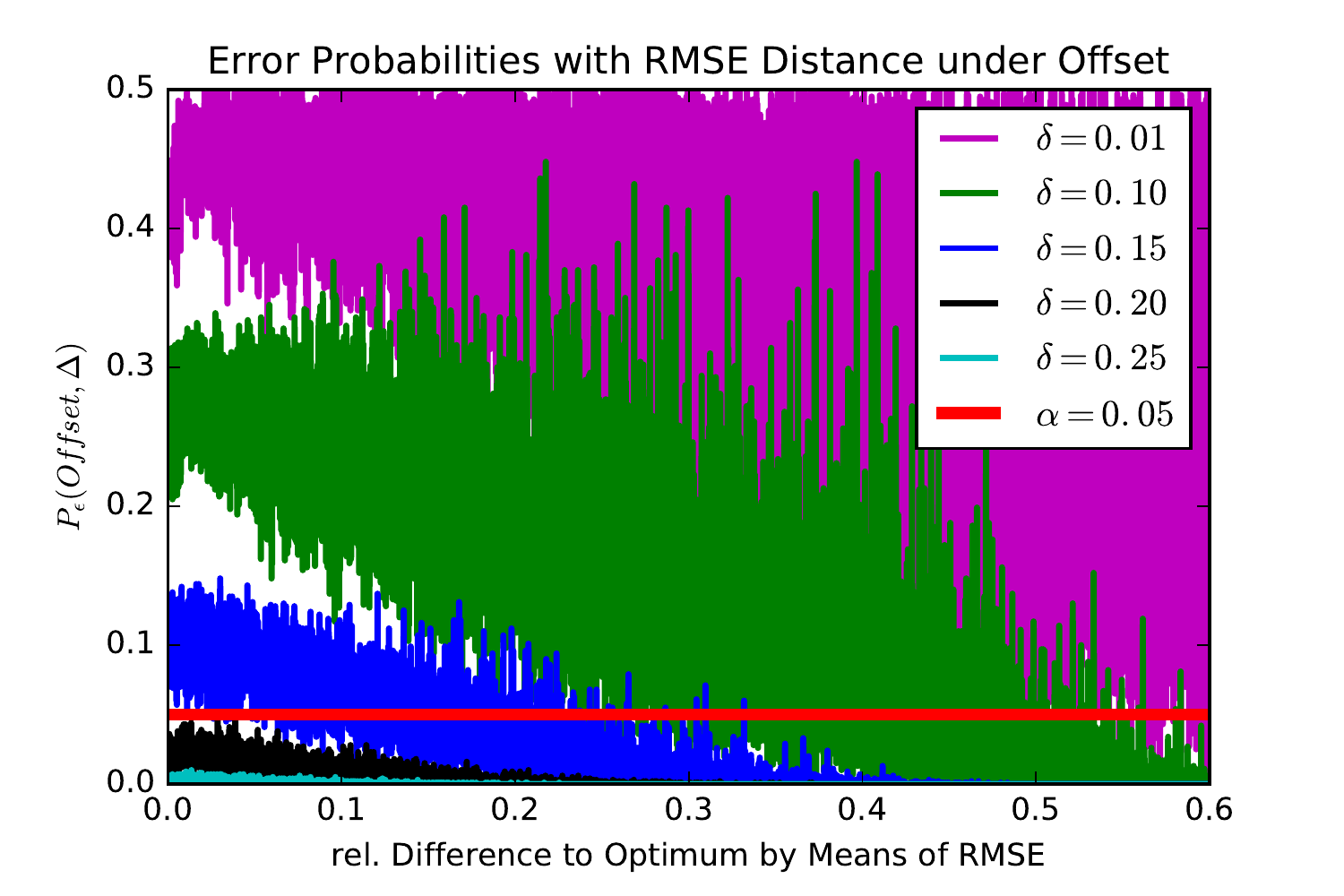}
        \caption{using relative differences of the RMSE as measure of distinction}
    \end{subfigure}
    \caption{Error probabilities for evaluating two suboptimal recommender systems for several distinctions under noise offset}
    \label{fig:RMSEDist}
\end{figure}
Figure \ref{fig:NoiseDist} depicts the family of curves mapping the noise offset to the corresponding error probabilities.
The offset represents background noise and is a measure of the deviation from the best possible recommender system, i.e. the larger the offset, the worse the prediction quality of the recommender system. The colours encode the relative difference $\Delta$ of two recommender systems among each other. For the green curve (representing 10\% noise of difference), an $x$-value of $0.15$ means that RS1 has a noise of 15\% whereas RS2 has a noise of 25\%. The corresponding $y$-value indicates the error probability for ranking both of these recommender systems using the point paradigm. 
It is apparent from this Figure, that two recommender systems can not be brought into a ranking order without considerable error probability if their relative difference is below 15\% , regardless of their fundamental prediction quality.
Figure \ref{fig:NoiseDist} also reveals that only for noise differences of more than 20\%, two different recommender systems can be resolved starting from a certain quality. As a result, we recognise the following: 
The better a systems becomes, the more improvement does a revision need in order to be detected with statistical evidence.

\item[Simulation 6:] 
In order to make our results more tangible and comparable to current competitions (e.g. the Netflix Prize), we define the RMSE difference as the relative difference in the expectation values of both distributions for this difference uses to be the best estimation for an infinitely repeated rating scenario. We rerun the last simulation, but now determine the RMSE distances by using adaptive noise: We only add so much noise until we reach the desired RMSE difference. Then we compare the error probabilities by means of those RMSE distances.

For the RMSE distances, a similar result is obtained as under simulation 5 (see Figure \ref{fig:RMSEDist}). 
Two recommender systems with a difference of 10\% in terms of RMSE must deviate more than 40\% from the optimum to be significantly be distinguished. In reverse interpretation, if the closeness of two recommender systems to the theoretical optimum (i.e. the offset) remains unknown - which is probably always the case in real life assessment - then both systems would only be distinguishable with statistical evidence, if they differ at least 20\% in terms of the RMSE (since only the 20\%-curve is below the 5\%-mark for any offset). 
\end{description}

\subsection*{Human Accuracy Metrics}
At this point, we investigate the resolution properties of two recommender systems by means of the sRMSE.
This is performed by a hypothesis test as described in Section 3 and considering only significant deviations from the rejection range to compute an RMSE. As a result, the sRMSE could theoretically distinguish between two recommender systems even with less deviations.
\begin{description}
\item[Simulation 7:] In practice, the hypothesis test is performed by constructing a symmetric interval around the predictor 
$\pi_\nu$ within the rating distribution of $R_\nu$ (step-range $10^{-3}$) until the density's area over this interval add to $0.95$. All values in this interval do not represent any significant deviations and are not taken into account in the sRMSE. We therefore generate pseudo-random numbers according to the distribution of $R_\nu$ until we have $\tau=10^6 $ values in the rejection range and use these to compute the sRMSE distributions. For these density functions, we now repeat the procedure from simulation 4.

\begin{figure}[t]
    \centering
        \includegraphics[width=.49\textwidth]{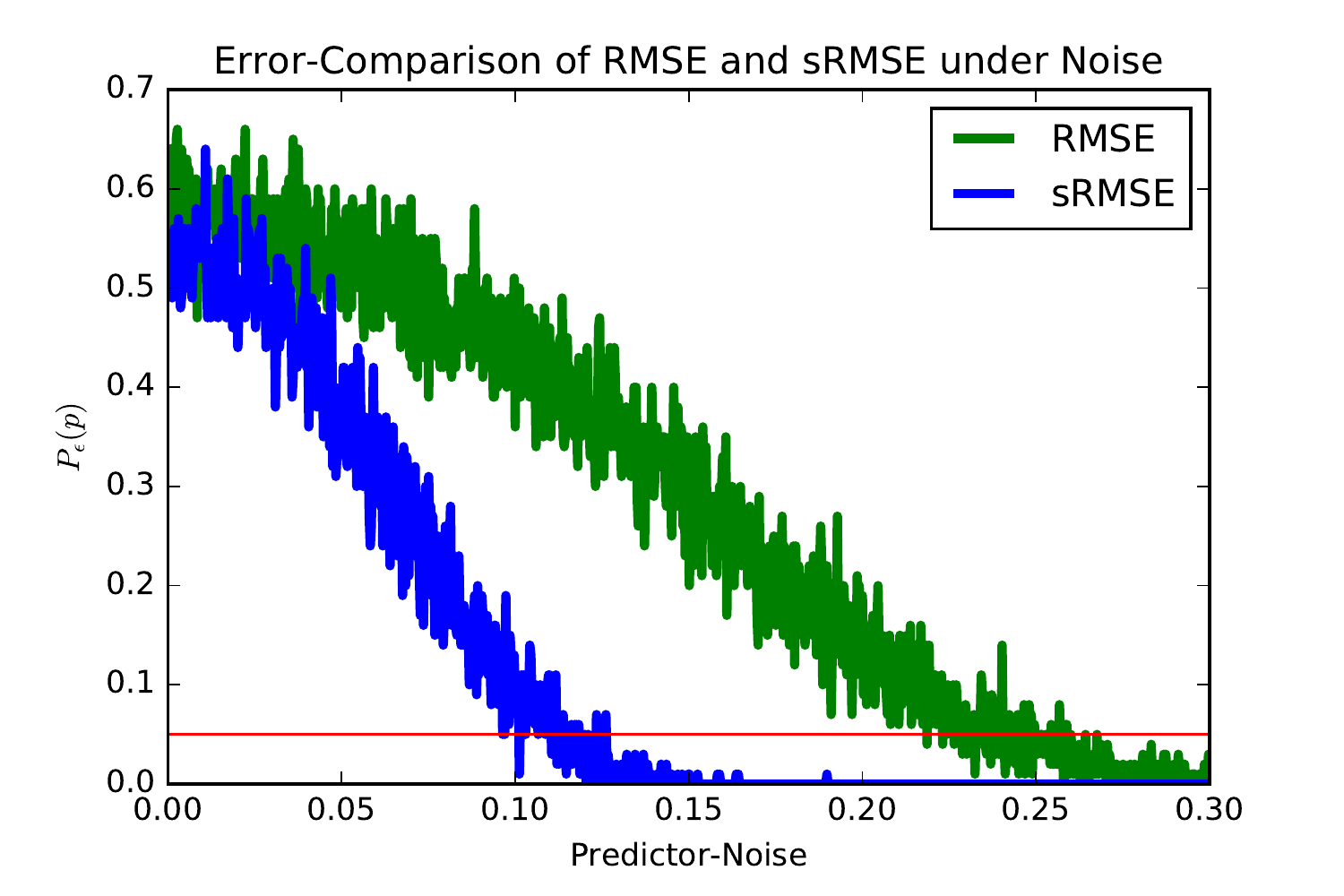} 
       \caption{A comparison of the RMSE and the sRMSE under artificial noise}
    \label{fig:sRMSEerr}
\end{figure}

The results are depicted in Figure \ref{fig:sRMSEerr}. 
Here we see error curves under noise in the form of a comparison of RMSE and sRMSE. 
It can be seen that the sRMSE grants substantially faster distinguishability from an optimum with statistically evidence than the traditional RMSE. Using this metric, a recommender system can already be distinguished from a theoretical optimum with 10\% of noise whereas the RMSE would probably need more than 20\%. A repetition of simulation 5 and 6 leads to equivalent results. This proves the better distinguishing features of the sRMSE as predicted by theory. 
\end{description}

\section{Discussion}
The lessons learned so far can be summarised as follows:
\begin{enumerate}
\item Due to the blur of the RMSE, an ordering relation is sometimes very difficult to define; we can only give probabilities for the existence of a particular order relation: The probability $P_\varepsilon:=P(R1>R2 \vert \mathbb{E}(R1)<\mathbb{E}(R2))$ for making an error when following the point-paradigm has proven to be an intuitive and very good metric. It correlates positively with the overlap area of two RMSE distributions and is therefore a good measure for the differentiation or resolution of two recommender systems and also serves as p-value for hypothesis testing.
\item A recommender system is only to be significantly distinguished from an optimum if it differs by more than 21 to 24\% in terms of noise. Below this limit, it cannot be distinguished with evidence.
\item The distinguishability of two recommender systems is not dependent solely on its (noise) difference, but also on their basic quality, that is, from their distance to a theoretical optimum. The worse two recommender systems predict, the less they have to differ in order to be distinguished evidently and vice versa.
\item Methods for collecting uncertainty information are yet to imprecise; the parameters of the rating distributions have such wide confidence intervals, that specifying RMSE densities is not reliable. We need between 500 and 1000 re-ratings for statements that exclude the worst case and about 2000 re-ratings for stable statements. The method of re-rating-proceeding as described in \cite{RateAgain} must therefore be improved.
\end{enumerate}

The most notable results are 2 and 3, since they show a natural limit for the resolution of evaluation metrics (which is also always present in the point paradigm but can not be made accessible). Result (2) implies the existence of an equivalence class of optimal recommenders because all recommender systems below a certain RMSE value are no longer to be distinguished from the optimum. Result (3) generalises this fact and raises the fundamental question of assessment evidence.
On the basis of our results, the suggested solution of using the sRMSE has proven to be quite fruitful for evaluating prediction quality. In the our simulations, the sRMSE outperformed the traditional RMSE by far, i.e. the resolution capability for two recommender systems was doubled. 

The implications of these findings for future assessment scenarios can be demonstrated by recent recommendation competitions like Netflix Prize or Movie Lens. In this contribution we will demonstrate our findings by example of the Netflix Prize competition \cite{netflixrules} for movie recommenders, specifically its publicly available ranking list \cite{netflixleaderboard}. 

Although the necessary user uncertainty information is not available for any of the recent competitions, we may consider the results of our case study (with trailer ratings similar to movie ratings at Netflix) for constructing realistic, indicative examples of possible advanced assessment interpretation under human uncertainty. 
In doing so we have to compare the RMSE scores of the best algorithms as well as the relative difference of these scores in table \ref{tab:Leader}. For a differentiated comparison, we have to distinct two cases:

{\bf Case 1:} Assuming that the Netflix in-house recommender Cinematch is already very close to the optimum (Offset $< 0.3$), the choice of the winner would be subject to a high probability of error up to 30\% that is, in one of three cases a serious mistake has been made. From Figure \ref{fig:RMSEDist} it is clear that a difference of only 10\% is not sufficient to distinguish a new recommender system from Cinematch. To be on the safe side, i.e. to allow significant distinctions for any basic quality (offset), a relative difference of at least 20\% would have been required, which is twice as much as what was done. So, if Cinematch is a very good recommender system, then there would probably no statistical evidence for the winner algorithm to be better indeed.

{\bf Case 2:} Assuming that the relative difference of 10 \% has very likely led to a significant distinction, then the winner algorithm must have an offset of at least 35\% according to Figure \ref{fig:RMSEDist}. This also implies that another recommender system must again differ by at least 10\%  from the winner in order to evidently reject their equality. In Table \ref{tab:Leader}, we see the 12 best recommender systems algorithms of the Netflix Prize and recognise that all these algorithms differ by less than 1\% from the winner. Accordingly, there are plenty of other algorithms whose equivalence to the winning algorithm can not be evidently rejected.
\begin{table}[t]
\centering
\renewcommand{\arraystretch}{1.2}
\small
\begin{tabular}{|l|p{1cm}|p{1.5cm}|}
\hline
Recommender System                      &       RMSE            &       Diff. to Winner \\ \hline
BellKor's Pragmatic Chaos       &       0.8567                          &       0.00    \%\\
The Ensemble                                            &       0.8567                          &       0.00            \% \\
Grand Prize Team                                &       0.8582                          &       0.17            \%      \\
...                                                     & ...                   & ... \\
BellKor                                                                         &       0.8624                          &       0.66             \% \\ \hline
Cinematch                                                       &       0.9525                          &       10.06   \% \\ \hline
\end{tabular} 
\caption{Best algorithms of the Netflix Prize (see \cite{netflixleaderboard})}
\label{tab:Leader}
\end{table}
In summary - if our findings may apply to the scenario of Netflix Prize - it can be assumed that the winner is perhaps either not really distinguishable from Cinematch or that many other algorithms are perhaps not distinguishable from the winner. Both cases indicate some difficulty in this evaluation and reveals more complexity in prediction quality assessment as commonly believed. A possible solution to this problem is provided by the sRMSE which could have been computed in this scenario if uncertainty information were available.

\section{Conclusion and Future Work}
In this contribution we consider recommender systems and their assessment by means of the RMSE. It has been shown that errors can be committed if human uncertainty is not included in prediction quality assessment. For example, the assessment of the Netflix Prize appears to be more complex according to the findings of our research as there is no statistical evidence for the decisions that have been made. 
It can be assumed that similar influences might also be observed for other metrics considering uncertain inputs in their computations, such as ratings and browsing behaviour. For example, the results presented here could be reproduced in equivalent form for the metrics average absolute deviation and mean signed deviation. Similar influences might be found not only in recommender systems, but also anywhere in predictive data mining where human behaviour is to be analysed. We were therefore able to provide initial indications that human uncertainty may have a striking influence on the predictive data mining and thus on all the areas that build upon it. 
On this basis, further research may lead into various directions:
For \textbf{theoretical research}, the overall goal is to develop a complete mathematical model of human uncertainty providing large connectivity for practical applications.
For \textbf{practical research} it would be quite profitable to assimilate technical approaches and sensitising them for human uncertainty. This could be done by developing a bayesian prediction models with informative priors based on advanced experiments.

\nocite{3DBenchmark, RateAgain, noise1, noise2}
\nocite{Kubat,DecMaking,ForschMeth,LatentModels,OneAndDone,PredUncer}

\bibliographystyle{ACM-Reference-Format}
\bibliography{Literature} 

\end{document}